\documentclass[11pt]{article}

\def\CM{\cal M}

\begin{document}

\date{}
\title{\textbf{Even and odd geometries on  supermanifolds}}
\author{\textsc{M.~Asorey}${}^{a}$\thanks{E-mail: asorey@saturno.unizar.es} and
\textsc{P.M.~Lavrov}${}^{b}$\thanks{E-mail: lavrov@tspu.edu.ru}\\
\\${}^a$\textit{Departamento de F\'{\i}sica Te\'{o}rica,}
\\\textit{Facultad de Ciencias Universidad de Zaragoza,}
\\\textit{50009 Zaragoza, Spain}\\
\\${}^b$\textit{Department of Mathematical Analysis,}
\\\textit{ Tomsk State Pedagogical University,}
\\\textit{ Tomsk 634041, Russia}}
\maketitle

\begin{quotation}
We analyze from a general perspective all possible supersymmetric generalizations  of symplectic and metric structures on smooth manifolds. There are  two different types of structures according  to the even/odd character of the corresponding quadratic tensors. In general we can have even/odd symplectic supermanifolds, Fedosov supermanifolds and    Riemannian supermanifolds. 
The geometry of even Fedosov supermanifolds is strongly constrained and has to be  flat. In the odd case,  the  scalar curvature  is only constrained by Bianchi identities. However, we show  that  odd Riemannian supermanifolds  can only have constant scalar curvature. We also point out that the supersymmetric generalizations of AdS space do not exist in the odd case.
\end{quotation}

\vspace{.5cm}

\section{Introduction}

The two main quadratic geometrical structures of smooth manifolds which play a significant role 
in classical and quantum physics  are  Riemannian metrics and symplectic
forms. Riemannian geometry is  not only  basic for  the formulation of general relativity 
 but also for  the very formulation
of gauge field theories. The symplectic structure  provides the geometrical framework 
for classical mechanics (see, e.g. \cite{Ar}) and  field theories
\cite{agm}.  The Fedosov method
of quantization  by deformation \cite{F} is also formulated in terms of  symplectic
structures and symplectic connections (the so-called Fedosov manifolds \cite{fm}). 
The introduction of the concept of  supermanifold by  Berezin \cite{Ber} (see also \cite{Leites, manin})
opened new perspectives for  geometrical approaches of supergravity and quantization of
gauge theories  \cite{bv, geom,btgl}. In summary, the geometry of manifolds and supermanifolds 
percolates all  fundamental physical theories.

In this note we address the classification of  possible extensions of  symplectic
and metric structures to supermanifolds in terms  of
graded symmetric and antisymmetric second-order tensor fields. 
The cases of even and odd  symplectic  and Riemannian supermanifolds
are analyzed in some detail.   Graded  non-degenerate Poisson
supermanifolds are described by symplectic supermanifolds that if
equipped with a symmetric symplectic connection  become graded Fedosov 
supermanifolds. The even case corresponds to a straightforward  generalization 
of Fedosov manifold \cite{fm} where the  scalar curvature vanishes as  
for standard  Fedosov manifolds. Graded metric supermanifolds 
equipped with the unique compatible symmetric connection also 
correspond to graded Riemannian supermanifold. The scalar curvature is  non trivial, 
in general, for  odd Riemannian  and  Fedosov supermanifolds, but in the first case 
it must always be constant. There is a supersymmetric generalization of AdS space but it is 
trivial in the odd case.

The paper is organized as follows.  In Sect.~2, we consider scalar structures
which can be used for the construction of symplectic and metric
supermanifolds.  The properties of   symmetric affine connections on supermanifolds
and  their curvature tensors are analyzed in Sect.~3.  In
Sect.~4, we introduce  the concepts of even and odd Fedosov
supermanifolds and  even and odd Riemannian supermanifolds are 
analyzed in Sect.~5.  Finally,   we  convey the main results in Sect.~6.
We use the condensed notation suggested by DeWitt \cite{DeWitt} and
definitions and notations adopted in \cite{al}. 

\section{Scalar Fields}
 
Let ${\cal M}$ be a supermanifold with a dimension $dim {\cal M}=N$ and 
 $\{x^i\}, \epsilon(x^i)=\epsilon_i$ a local
system of coordinates on in the vicinity of a
point $p\in \CM$. Let us consider now the most general  scalar structures on
supermanifolds which can be defined in terms of graded second-rank symmetric and
antisymmetric tensor fields.

In general, there exist eight types of second rank tensor fields
with the required symmetry properties
\begin{eqnarray}
\label{tfant1}
&&\omega^{ij}=-(-1)^{\epsilon_i\epsilon_j}\omega^{ji},\quad
\epsilon(\omega^{ij})=
\epsilon(\omega)+\epsilon_i+\epsilon_j,\\
\label{tfsyt1}
&&\Omega^{ij}=(-1)^{\epsilon_i\epsilon_j}\Omega^{ji},\;\;\;\quad
\epsilon(\Omega^{ij})=
\epsilon(\Omega)+\epsilon_i+\epsilon_j,\\
\label{tfant2}
&&E_{ij}=-(-1)^{\epsilon_i\epsilon_j}E_{ji},\quad \epsilon(E_{ij})=
\epsilon(E)+\epsilon_i+\epsilon_j,\\
\label{tfsyt2}
&&g_{ij}=(-1)^{\epsilon_i\epsilon_j}g_{ji}, \;\;\;\;\;\;\quad
\epsilon(g_{ij})=
\epsilon(g)+\epsilon_i+\epsilon_j.
\end{eqnarray}
Using these tensor fields (\ref{tfant1})-(\ref{tfsyt2})
it is not difficult to
built eight scalar structures on a supermanifold:
\begin{eqnarray}
\label{Pst}
\{A,B\}&=&\frac{\partial_r A}{\partial x^i}(-1)^{\epsilon_i\epsilon(\omega)}
\omega^{ij}\frac{\partial B}{\partial x^j},\quad \epsilon(\{A,B\})=
\epsilon(\omega)+\epsilon(A)+\epsilon(B),\\
\label{Ant}
(A,B)&=&\frac{\partial_r A}{\partial x^i}(-1)^{\epsilon_i\epsilon(\Omega)}
\Omega^{ij}\frac{\partial B}{\partial x^j},\quad \epsilon((A,B))=
\epsilon(\Omega)+\epsilon(A)+\epsilon(B),\\
\label{E}
E&=&E_{ij}dx^j\land dx^i,\quad\quad\quad\;\;\;\;\;\;
\epsilon(E_{ij}dx^j\land dx^i)=\epsilon(E),\\
\label{g}
g&=&g_{ij}dx^j\;dx^i,\quad\quad\quad\quad\;\;\;\;\;\;
\epsilon(g_{ij}dx^j\;dx^i)=\epsilon(g),
\end{eqnarray}
where $A$ and $B$ are arbitrary superfunctions.

The bilinear operation $\{A,B\}$ (\ref{Pst}) obeys the following symmetry
property
\begin{eqnarray}
\label{Pstsym}
\{A,B\}=-(-1)^{\epsilon(\omega)+(\epsilon(A)+
\epsilon(\omega))(\epsilon(B)+\epsilon(\omega))}\{B,A\}
\end{eqnarray}
which  in the even case ($\epsilon(\omega)=0$) reduces to
\begin{eqnarray}
\label{Pstsyme}
\{A,B\}=-(-1)^{(\epsilon(A)\epsilon(B)}\{B,A\}
\end{eqnarray}
and in the odd case  ($\epsilon(\omega)=1$) to
\begin{eqnarray}
\label{Pstsymo}
\{A,B\}=(-1)^{(\epsilon(A)+1))(\epsilon(B)+1)}\{B,A\}.
\end{eqnarray}
On the other hand, the bilinear operation $(A,B)$ (\ref{Ant}) has the symmetry property
\begin{eqnarray}
\label{Antsym}
(A,B)=(-1)^{\epsilon(\omega)+(\epsilon(A)+
\epsilon(\omega))(\epsilon(B)+\epsilon(\omega))}(B,A)
\end{eqnarray}
which  in the even case ($\epsilon(\omega)=0$) reduces to
\begin{eqnarray}
\label{Antsyme}
(A,B)=(-1)^{\epsilon(A)\epsilon(B)}(B,A)
\end{eqnarray}
and in the odd case  ($\epsilon(\omega)=1$) to
\begin{eqnarray}
\label{Anttsymo}
(A,B)=-(-1)^{(\epsilon(A)+1))(\epsilon(B)+1)}(B,A).
\end{eqnarray}

One can easily check  that  in the even case ($\epsilon(\omega)=0$)
the bilinear operation $\{A,B\}$ satisfies
the Jacobi identity
\begin{eqnarray}
\label{PstJI}
 \{A,\{B,C\}\}(-1)^{\epsilon(A)(\epsilon(C)}+ \{C,\{A,B\}\}(-1)^{\epsilon(C)(\epsilon(B)}+
  \{B,\{C,A\}\}(-1)^{\epsilon(B)(\epsilon(A)}\equiv 0
\end{eqnarray}
if and only if $\omega$ satisfies 
\begin{eqnarray}
\label{PstJIw}
\omega^{ij}
\frac{\partial \omega^{kl}}{\partial x^j}
(-1)^{\epsilon_i\epsilon_l}+ 
\omega^{lj}\frac{\partial \omega^{ik}}{\partial x^j}
(-1)^{\epsilon_l\epsilon_k}+\omega^{kj}
\frac{\partial \omega^{li}}{\partial x^j}
(-1)^{\epsilon_k\epsilon_i}\equiv 0.
\end{eqnarray}
In the odd case there is no possibility of satisfying the Jacobi identity for 
the operation $\{A,B\}$.
 
On the contrary, the 
 Jacobi's identity for $(A,B)$ can be satisfied
\begin{eqnarray}
\label{AntJI}
(A,(B,C))(-1)^{(\epsilon(A)+1)(\epsilon(C)+1)}+
(C,(A,B))(-1)^{(\epsilon(C)+1)(\epsilon(B)+1)}+
(B,(C,A))(-1)^{(\epsilon(B)+1)(\epsilon(A)+1)} \equiv 0
\nonumber
\end{eqnarray}
if and only if  $\Omega$ is odd, $\epsilon(\Omega)=1$, and satisfies
\begin{eqnarray}
\label{AntJIW}
\Omega^{ij}
\frac{\partial \Omega^{kl}}{\partial x^j}
(-1)^{\epsilon_i(\epsilon_l+1)}+ \Omega^{lj}
\frac{\partial \Omega^{ik}}{\partial x^j}
(-1)^{\epsilon_l(\epsilon_k+1)}+
\Omega^{kj}
\frac{\partial \Omega^{li}}{\partial x^j}
(-1)^{\epsilon_k(\epsilon_i+1)}
\equiv 0. 
\end{eqnarray}
 Therefore, because of the  identities (\ref{PstJIw}) and
(\ref{AntJIW}), one can identify $\{A,B\}$
($\epsilon(\{A,B\})=\epsilon(A)+\epsilon(B)$) and $(A,B)$
($\epsilon((A,B))=\epsilon(A)+\epsilon(B)+1$) with the Poisson
bracket and the antibracket respectively.

It is also possible to  combine the Poisson bracket associated to $\omega$ and the antibracket into the
so-called graded Poisson bracket (see, for example, \cite{BB, Bering,CarF,Bering1})
in the following  bilinear operation
\begin{eqnarray}
\label{Pstg}
\{A,B\}_g&=&\frac{\partial_r A}{\partial x^i}
(-1)^{\epsilon_i\epsilon(\omega_g)}
\omega^{ij}_{g}\frac{\partial B}{\partial x^j},\quad
\omega^{ij}_{g}=-(-1)^{\epsilon(\omega_g+\epsilon_i\epsilon_j)}\omega^{ji}_{g},\\
\nonumber
&&\epsilon(\{A,B\}_g)=
\epsilon(\omega_g)+\epsilon(A)+\epsilon(B).
\label{omegacom}
\end{eqnarray}
From (\ref{omegacom}) it follows the  symmetry property
\begin{eqnarray}
\label{Pstsymg}
\{A,B\}_g=-(-1)^{(\epsilon(A)+
\epsilon(\omega_g))(\epsilon(B)+\epsilon(\omega_g))}\{B,A\}_g.
\end{eqnarray}
If the tensor fields $\omega^{ij}$ satisfy the identities
\begin{eqnarray}
\label{PstJIwg}
\omega^{ij}_g
\frac{\partial \omega^{kl}_g}{\partial x^j}
(-1)^{\epsilon_i(\epsilon_l+\epsilon(\omega_g))}+\omega^{lj}_g \frac{\partial \omega^{ik}_g}{\partial x^j}
(-1)^{\epsilon_l(\epsilon_k+\epsilon(\omega_g))}+\omega^{kj}_g\frac{\partial \omega^{li}_g}{\partial x^j}
(-1)^{\epsilon_k(\epsilon_i+\epsilon(\omega_g))} \equiv 0,
\end{eqnarray}
then $\{A,B\}_g$ satisfies the Jacobi identity
\begin{eqnarray}
\label{AntJIg} \{A,\{B,C\}_g\}_g(-1)^{\epsilon_g(A,B.C)}+
 \{C,\{A,B\}_g\}_g(-1)^{\epsilon_g(B,C,A)}
+\{B,\{C,A\}_g\}_g(-1)^{\epsilon_g(C,A,B)} \equiv 0
\end{eqnarray}
with $\epsilon_g(A,B,C)={(\epsilon(A)+\epsilon(\omega_g))(\epsilon(C)+\epsilon(\omega_g))}$
and plays the role of a graded Poisson bracket.

 A supermanifold $\CM$
equipped with a Poisson bracket is called a Poisson supermanifold,
$({\CM}, \{,\})$. Usually a manifold $\CM$ equipped with an non-degenerate
antibracket is called an antisymplectic supermanifold $({\CM}, (,))$ or,
sometimes, an odd Poisson supermanifold (see, for example,
\cite{CarF,Bering1}).

In Eq. (\ref{tfant2}) $E$ denotes a generic graded differential 2-form. If $E$ is closed
\begin{eqnarray}
\label{Eclos}
dE=E_{ij,k}dx^k\land dx^j\land dx^i=0
\end{eqnarray}
and non-degenerate, then it defines a graded (even or odd)
symplectic supermanifold $({\CM},E)$
\cite{Leites}.
In terms of tensor fields $E_{ij}$ the  condition  (\ref{Eclos}) can be expressed as
\begin{eqnarray}
\label{Eclos1}
E_{ij,k}(-1)^{\epsilon_i\epsilon_k}+E_{jk,i}(-1)^{\epsilon_j\epsilon_i}+
E_{ki,j}(-1)^{\epsilon_k\epsilon_j}=0, \quad E_{ij}=
-(-1)^{\epsilon_i\epsilon_j}E_{ji}
\end{eqnarray}
and in terms of inverse tensor fields $E^{ij}$ Eqs. (\ref{Eclos1})
can be rewritten in the form
\begin{eqnarray}
\label{Eclosivn}
E^{il}\frac{\partial E^{jk}}{\partial x^l}
(-1)^{\epsilon_i(\epsilon_k+\epsilon(E))}+
E^{kl}\frac{\partial E^{ij}}{\partial x^l}(-1)^{\epsilon_k(\epsilon_j+\epsilon(E))}+
E^{jl}\frac{\partial E^{ki}}{\partial x^l}
(-1)^{\epsilon_j(\epsilon_i+\epsilon(E))}=0,
\end{eqnarray}
where $E^{ij}=-(-1)^{\epsilon(E)+\epsilon_i\epsilon_j}E^{ji}$.
Identifying $E^{ij}$ with the tensor field $\omega^{ij}$ in (\ref{Pst}),
one gets in the even case ($\epsilon(E)=0$) the Poisson bracket for which the
Jacobi identity (\ref{PstJI}) follows from (\ref{Eclosivn}). Therefore, in
the even case there is one-to-one correspondence between  non-degenerate
Poisson supermanifolds and an even symplectic supermanifolds. 
In the odd case ($\epsilon(E)=1$), if we assume  $E^{ij}=\Omega^{ij}$ in  (\ref{Ant}) 
then $E^{ij}$ defines an antibracket for which the Jacobi identity (\ref{AntJIW}) 
follows from (\ref{Eclosivn}). Therefore  antisymplectic supermanifolds can be identified 
with  odd symplectic manifolds.

If the tensor field $g_{ij}$ in (\ref{g}) is non-degenerate, one has a
graded metric that can provide a  supermanifold $\CM$  with a
graded (even or odd) metric structure, giving rise to a Riemannian supermanifold $({{\CM}},g)$.
On the other hand, the inverse tensor field $g^{ij}$ also defines a bilinear operation 
with symmetry properties (\ref{Pstsymo}) or (\ref{Antsyme}) 
but it does not satisfy  the Jacobi identity.
\\

\section{Connections in Supermanifolds}

Let us consider a covariant derivative $\nabla$ (or an affine connection
$\Gamma$) on a supermanifold ${\cal M}$. In each local coordinate system $\{ x \}$
the covariant derivative $\nabla$ is described by its components $\nabla_i \,
(\epsilon(\nabla_i)= \epsilon_i)$, which are related to the 
components the affine connection $\Gamma$
$\Gamma^i_{\;\;jk},\; (\epsilon(\Gamma^i_{\;\;jk})=\epsilon_i+\epsilon_j+
\epsilon_k)$
by
\begin{eqnarray}
\label{Cris}
e^i\nabla_j=e^k\Gamma^i_{\;\;kj}(-1)^{\epsilon_k(\epsilon_i+1)},
\quad e_i\nabla_j=-e_k\Gamma^k_{\;\;ij}
\end{eqnarray}
where $\{e_i\}$ and $\{e^i\}$ are the associated bases of   the tangent 
$T\CM$ and cotangent $T^\ast \CM$
 spaces respectively.
The action of the covariant derivative on a tensor field
of any rank and type is given in terms of the tensor components,
the ordinary derivatives and the connection components 
(for details see \cite{al}). From here on, we shall consider only 
symmetric connections
\begin{eqnarray}
\label{Crisp} \Gamma^i_{\;jk}=
(-1)^{\epsilon_j\epsilon_k}\Gamma^i_{\;kj}.
\end{eqnarray}

The curvature tensor field $R^i_{\;\;mjk}$ is defined in terms  of the commutator of covariant
derivatives, $[\nabla_i,\nabla_j]=
\nabla_i\nabla_j-(-1)^{\epsilon_i\epsilon_j}\nabla_j\nabla_i$, whose action 
on a vector field $T^i$ is
\begin{eqnarray}
\label{Rie} T^i[\nabla_j,\nabla_k]=-(-1)^{\epsilon_m(\epsilon_i+1)}
T^mR^i_{\;\;mjk}.
\end{eqnarray}
The choice of factor in r.h.s (\ref{Rie}) is dictated by the
requirement that the contraction of tensor fields of types $(1,0)$ and
$(1,3)$ yield  a tensor field of type $(1,2)$. A straightforward
calculation yields
\begin{eqnarray}
\label{R} R^i_{\;\;mjk}=-\Gamma^i_{\;\;mj,k}+
\Gamma^i_{\;\;mk,j}(-1)^{\epsilon_j\epsilon_k}+
\Gamma^i_{\;\;jn}\Gamma^n_{\;\;mk}(-1)^{\epsilon_j\epsilon_m}-
\Gamma^i_{\;\;kn}\Gamma^n_{\;\;mj}
(-1)^{\epsilon_k(\epsilon_m+\epsilon_j)}.
\end{eqnarray}
The curvature tensor field has a generalized
antisymmetry,
\begin{eqnarray}
\label{Rsym}
R^i_{\;\;mjk}=-(-1)^{\epsilon_j\epsilon_k}R^i_{\;\;mkj}\,;
\end{eqnarray}
and  satisfies  the  Jacobi identity,
\begin{eqnarray}
\label{Rjac} (-1)^{\epsilon_m\epsilon_k}R^i_{\;\;mjk}
+(-1)^{\epsilon_j\epsilon_m}R^i_{\;\;jkm}
+(-1)^{\epsilon_k\epsilon_j}R^i_{\;\;kmj}\equiv 0\,.
\end{eqnarray}
Using the  Jacobi identity for the covariant derivatives,
\begin{eqnarray}
\label{}
[\nabla_i,[\nabla_j,\nabla_k]](-1)^{\epsilon_i\epsilon_k}+
[\nabla_k,[\nabla_i,\nabla_j]](-1)^{\epsilon_k\epsilon_j}+
[\nabla_j,[\nabla_k,\nabla_i]](-1)^{\epsilon_i\epsilon_j}\equiv
0\,,
\end{eqnarray}
one obtains the  Bianchi identity,
\begin{eqnarray}
\label{BI} (-1)^{\epsilon_i\epsilon_j}R^n_{\;\;mjk;i}
+(-1)^{\epsilon_i\epsilon_k}R^n_{\;\;mij;k}
+(-1)^{\epsilon_k\epsilon_j}R^n_{\;\;mki;j}\equiv 0\,,
\end{eqnarray}
with the notation $R^n_{\;\;mjk;i}:\,=R^n_{\;\;mjk}\nabla_i$.
\\

\section{Symplectic supermanifolds}

Let us consider a symplectic supermanifold $({\CM},\omega)$, 
i.e. a supermanifold $\CM$  with 
a closed non-degenerate graded differential 2-form $\omega$
\begin{eqnarray}
 \label{omega1}
\omega=\omega_{ij}dx^j\wedge dx^i,\quad \omega_{ij}=
-(-1)^{\epsilon_i\epsilon_j}\omega_{ji}.
\end{eqnarray}
The closure condition of $\omega$, $d\omega=0$,
can be rewritten as
\begin{eqnarray}
\label{domega} 
\omega_{ij,k}(-1)^{\epsilon_i\epsilon_k}+\omega_{jk,i}(-1)^{\epsilon_i\epsilon_j}+
\omega_{ki,j}(-1)^{\epsilon_j\epsilon_k}=0
\end{eqnarray}
in terms of the inverse tensor field $\omega^{ij}$
\begin{eqnarray}
\label{omegainvsy}  
\omega^{ij}=-(-1)^{\epsilon(\omega)+\epsilon_i\epsilon_j}\omega^{ji}
\end{eqnarray}
and do coincide with identities (\ref{PstJIwg}). It means that in the even case ($\epsilon(\omega)=0$) $\omega^{ij}$ defines a nondegenerate Poisson bracket 
while in the odd case ($\epsilon(\omega)=1$)  it defines an antibracket. Therefore
in the even case there is a one-to-one correspondence between even symplectic supermanifolds and nondegenerate Poisson supermanifold. In the odd case any antisymplectic supermanifold is nothing but an odd symplectic supermanifold. 

Let $\Gamma$ be a symmetric connection  of a symplectic supermanifold $({\CM},\omega)$. 
The corresponding covariant derivative $\nabla$ has to verify the compatibility condition
 $\omega\nabla=0$  with the 
 symplectic structure $\omega$. 
In each local coordinate system
$\{x^i\}$ 
 the   compatibility condition can be expressed as
\begin{eqnarray}
\label{covomiv} \omega_{ij}\nabla_k=\omega_{ij,k}-\Gamma_{ijk}+
\Gamma_{jik}(-1)^{\epsilon_i\epsilon_j}=0,\quad 
\omega_{ij}=-(-1)^{\epsilon_i\epsilon_j}\omega_{ji}
\end{eqnarray}
in terms of the  components $\Gamma^i_{\;jk}$ ($\nabla_i$) of   the symplectic connection $\nabla$,
where we use the notation
\begin{eqnarray}
\label{G} \Gamma_{ijk}=\omega_{in}\Gamma^n_{\;\;jk},\quad
\epsilon(\Gamma_{ijk})=\epsilon(\omega)+
\epsilon_i+\epsilon_j+\epsilon_k\,.
\end{eqnarray}

A symplectic supermanifold $({\CM},\omega)$
equipped with a symmetric symplectic connection $\Gamma$ is called
a Fedosov supermanifold $({\CM},\omega,\Gamma)$.

Let us consider now  curvature tensor $R_{ijkl}$ of a  symplectic connection
\begin{eqnarray}
\label{Rs} R_{ijkl}=\omega_{in}R^n_{\;\;jkl},\quad
\epsilon(R_{ijkl})=\epsilon(\omega)+\epsilon_i+
\epsilon_j+\epsilon_k+\epsilon_l,
\end{eqnarray}
where  $R^n_{\;\;jkl}$ is defined in  (\ref{R}).
This tensor has the following symmetry properties 
\begin{eqnarray}
\label{Rans} R_{ijkl}=-(-1)^{\epsilon_k\epsilon_l}R_{ijlk},\quad
R_{ijkl}=(-1)^{\epsilon_i\epsilon_j}R_{jikl}
\end{eqnarray}
and satisfies the identity
\begin{eqnarray}
\label{Rjac2}
R_{ijkl}
+(-1)^{\epsilon_l(\epsilon_i+\epsilon_k+\epsilon_j)}R_{lijk}
+(-1)^{(\epsilon_k+\epsilon_l)(\epsilon_i+\epsilon_j)}
R_{klij}+
(-1)^{\epsilon_i(\epsilon_j+\epsilon_l+\epsilon_k)}R_{jkli}=0.
\end{eqnarray}
The last statement  can be derived from  the Jacobi identity
\begin{eqnarray}
\label{Rjac0} (-1)^{\epsilon_j\epsilon_l}R_{ijkl}
+(-1)^{\epsilon_l\epsilon_k}R_{iljk}
+(-1)^{\epsilon_k\epsilon_j}R_{iklj}=0\,.
\end{eqnarray}
together with a cyclic change of  indices
\cite{lr}. The identity (\ref{Rjac2}) involves different components of the
curvature tensor with cyclic permutation of all indices, but the
sign factors depend  on the Grassmann parities of the indices and do
not follow  a cyclic permutation rule, similar
to that of Jacobi identity, but are defined by the permutation of the
indices that maps a given set into the original one.

From  the curvature tensor, $R_{ijkl}$, and the inverse tensor
field $\omega^{ij}$  of the symplectic structure $\omega_{ij}$,
 one can construct the only canonical tensor field of type $(0,2)$,
\begin{eqnarray}
 \label{R2} K_{ij}= \omega^{kn}R_{nikj}
 (-1)^{\epsilon_i\epsilon_k+(\epsilon(\omega)+1)(\epsilon_k+\epsilon_n)}
 \;=\;R^k_{\;\;ikj}\;(-1)^{\epsilon_k(\epsilon_i+1)},\quad
 \epsilon(K_{ij})=\epsilon_i+\epsilon_j.
\end{eqnarray}
This tensor $K_{ij}$ is the Ricci tensor and satisfies  the relations \cite{gl}
\begin{eqnarray}
\label{Rl3} [1+(-1)^{\epsilon(\omega)}](K_{ij}-(-1)^{\epsilon_i\epsilon_j}K_{ji})=0.
\end{eqnarray}
 In the even case  $K_{ij}$ is symmetric  
whereas  in the odd case there are not restrictions on its (generalized)
symmetry properties. 

Now, one can define the scalar curvature tensor $K$ by the formula
\begin{eqnarray}
\label{Rsc} K=\omega^{ji}K_{ij}(-1)^{\epsilon_i+\epsilon_j}=
\omega^{ji}\omega^{kn}R_{nikj}
(-1)^{\epsilon_i+\epsilon_j+\epsilon_i\epsilon_k+
(\epsilon(\omega)+1)(\epsilon_k+\epsilon_n)}.
\end{eqnarray}
From the symmetry properties of $R_{ijkl}$, it 
follows that
\begin{eqnarray}
\label{Rsc1} [1+(-1)^{\epsilon(\omega)}]K=0,
\end{eqnarray}
which  proves  that as in the case of Fedosov manifolds \cite{fm} the scalar curvature $K$
vanishes. 


However, for odd Fedosov supermanifolds $K$  is, in general, 
non-vanishing. This fact was quite recently used in Ref. \cite{BB} to  generalize
the BV formalism \cite{bv}.

Let us consider the Bianchi identity (\ref{BI}) in the form
\begin{eqnarray}
\label{BIFSM} 
R^n_{\;\;mij;k}
-R^n_{\;\;mik;j}(-1)^{\epsilon_k\epsilon_j}
+R^n_{\;\;mjk;i}(-1)^{\epsilon_i(\epsilon_j+\epsilon_k)}\equiv 0\,.
\end{eqnarray}
Contracting indices $i$ and $n$  with the help of (\ref{R2}) we obtain 
\begin{eqnarray}
\label{BIFSM1} 
K_{mj;k}
-K_{mk;j}(-1)^{\epsilon_k\epsilon_j}
+R^n_{\;\;mjk;n}(-1)^{\epsilon_n(\epsilon_m+\epsilon_j+\epsilon_k+1)}\equiv 0\,.
\end{eqnarray}
Now using  the relations
 \begin{eqnarray}
&& K^{i}_{\;\;j}=\omega^{ik}K_{kj}(-1)^{\epsilon_k},
K^{i}_{\;\;j;m}=\omega^{ik}K_{kj;m}(-1)^{\epsilon_k}
\\
&& 
K^{i}_{\;j;i}(-1)^{\epsilon_i(\epsilon_j+1)}= \omega^{ik}K_{kj;i}(-1)^{\epsilon_k+\epsilon_i(\epsilon_j+1)},
\end{eqnarray}
 it follows that 
\begin{eqnarray}
\label{Riccisc1} 
K_{,i}=[1-(-1)^{\epsilon(\omega)}]K^{j}_{\;\;i;j}
(-1)^{\epsilon_j(\epsilon_i+1)}.
\end{eqnarray}
In the odd case this implies that
\begin{eqnarray}
\label{Riccisc} 
K_{,i}=2K^{j}_{\;\;i;j}
(-1)^{\epsilon_j(\epsilon_i+1)}.
\end{eqnarray}
In the even case $K_{,i}=0$ but  in that case  the relation (\ref{Riccisc1}) does not provides 
any  new information because in this case $K=0$.
\\

\section{Riemannian supermanifolds}

Let     ${\CM}$ be a supermanifold  equipped both with a metric
structure $g$
\begin{eqnarray}
\label{gform} g=g_{ij}\;dx^j dx^i,\quad
g_{ij}=(-1)^{\epsilon_i\epsilon_j}g_{ji}, \quad
\epsilon(g_{ij})=\epsilon(g)+\epsilon_i+\epsilon_j\;,
\end{eqnarray}
and  a  symmetric connection  $\Delta$ with a covariant derivative
$\nabla$  compatible with the super-Riemannian metric  $g$
\begin{eqnarray}
\label{covgdiv} g_{ij}\nabla_k=g_{ij,k}-g_{im}\Delta^m_{\;\;jk}-
g_{jm}\Delta^m_{\;\;ik}(-1)^{\epsilon_i\epsilon_j}=0.
\end{eqnarray}

It is easy to show that as in the case of Riemannian geometry  there exists the unique symmetric
connection $\Delta^i_{\;jk}$ which is compatible with a given metric
structure. Indeed, proceeding   in the same way as in the usual
Riemannian geometry one obtains the generalization of celebrated Christoffel  formula
for the connection in supersymmetric case \cite{al}
\begin{eqnarray}\label{gDelta}
\Delta^l_{\;ki}=\frac{1}{2}g^{lj}\Big(g_{ij,k}
(-1)^{\epsilon_k\epsilon_i} +g_{jk,i}(-1)^{\epsilon_i\epsilon_j}-
g_{ki,j}(-1)^{\epsilon_k\epsilon_j}\Big)(-1)^
{\epsilon_j\epsilon_i+\epsilon_j+\epsilon(g)(\epsilon_j+\epsilon_l)}.
\end{eqnarray}
It is straightforward to show  that the symbols $\Delta^l_{\;ki}$ in
(\ref{gDelta}) are transformed according  with transformation
laws  for connections. A metric supermanifold $({{\CM}},g)$
equipped with a (even or odd) symmetric connection $\Delta$
compatible with a given metric structure $g$ is called  a
(even or odd) Riemannian supermanifold $({{\CM}},g,\Delta)$.

The curvature tensor of the connection $\Delta$ is
(\ref{gDelta})
\begin{eqnarray}
\label{Rs} {\cal R}_{ijkl}=g_{in}{\cal R}^n_{\;jkl},\quad
\epsilon({\cal R}_{ijkl})=\epsilon(g)+\epsilon_i+
\epsilon_j+\epsilon_k+\epsilon_l,
\end{eqnarray}
where ${\cal R}^n_{\;\;jkl}$ is given by (\ref{R}) by replacing
 $\Gamma^i_{\;jk}$ for $\Delta^i_{\;jk}$.
The curvature tensor has the following symmetry properties \cite{al}
\begin{eqnarray}
\label{Rans1} {\cal R}_{ijkl}=-(-1)^{\epsilon_k\epsilon_l}{\cal R}_{ijlk},
\quad
{\cal R}_{ijkl}=-(-1)^{\epsilon_i\epsilon_j}{\cal R}_{jikl}, \quad
{\cal R}_{ijkl}={\cal R}_{klij}(-1)^{(\epsilon_i+\epsilon_j)
(\epsilon_k+\epsilon_l)}.
\end{eqnarray}

From the  curvature tensor ${\cal R}_{ijkl}$ and the inverse tensor field $g^{ij}$ of the metric $g_{ij}$
defined by
\begin{eqnarray}
\label{ginv}
g^{ij}=(-1)^{\epsilon(g)+\epsilon_i\epsilon_j}g^{ji},\quad 
\epsilon(g^{ij})=\epsilon(g)+\epsilon_i+\epsilon_j,
\end{eqnarray}
one can define the only independent tensor field of type $(0,2)$:
\begin{eqnarray}
\label{3tf2}
&&{\cal R}_{ij}={\cal R}^k_{\;\;ikj}(-1)^{\epsilon_k(\epsilon_i+1)}=
g^{kn}{\cal R}_{nikj}
(-1)^{(\epsilon_k+\epsilon_n)(\epsilon(g)+1)+\epsilon_i\epsilon_k},\\
\nonumber
&&\epsilon({\cal R}_{ij})=\epsilon_i+\epsilon_j.
\end{eqnarray}
It  is the generalized Ricci tensor 
which  obeys  the symmetry 
\begin{eqnarray}
\label{sp2r}
{\cal R}_{ij}=(-1)^{\epsilon(g)+\epsilon_i\epsilon_j}{\cal R}_{ji}.
\end{eqnarray}
 
A further contraction between the metric and Ricci tensors
defines  the scalar curvature
\begin{eqnarray}
\label{Scalcur} {\cal R} = g^{ji}{\cal
R}_{ij}\;(-1)^{\epsilon_i+\epsilon_j},\quad \epsilon({\cal
R})=\epsilon(g)
\end{eqnarray}
which, in general, is non vanishing. Notice that for an odd
metric structure the scalar curvature tensor squared is identically
equal to zero, ${\cal R}^2=0$.

Let us consider now relations which follow from the Bianchi identity (\ref{BI}).
Repeating all arguments given in the end of previous Section one can derive  the following 
relation between the  scalar curvature  and the Ricci tensor
\begin{eqnarray}
\label{RicciscR1} 
{\cal R}_{,i}=[1+(-1)^{\epsilon(g)}]{\cal R}^{j}_{\;\;i;j}
(-1)^{\epsilon_j(\epsilon_i+1)}.
\end{eqnarray}
In the even case we have
\begin{eqnarray}
\label{Riccisc} 
{\cal R}_{,i}=2{\cal R}^{j}_{\;\;i;j}
(-1)^{\epsilon_j(\epsilon_i+1)},
\end{eqnarray}
which is a supersymmetric generalization of the well known relation of Riemannian geometry \cite{Eisenhart}.
In the odd case ${\cal R}_{,i}=0$ and the relation (\ref{RicciscR1}) implies that  ${\cal R}=$const. 

Therefore, odd Riemann supermanifolds can only have constant scalar curvature ${\cal R}=$ const.


It is well known that  special types of Riemannian manifolds play an important role
in modern quantum field theory. In particular, a consistent formulation of higher spin field theories is possible on AdS space (see, for example \cite{hsft}). In this case the curvature, Ricci  and scalar curvature tensors have the form
\begin{eqnarray}
\label{ads} 
{\cal R}_{ijkl}=R(g_{ik}g_{jl}-g_{il}g_{jk}),\quad {\cal R}_{ij}=(N-1)Rg_{ij},\quad {\cal R}= N(N-1)R,
\end{eqnarray}
where $N$ is the dimension of the Riemannian manifold $\CM$  with a metric tensor 
$g_{ij}$ and $R$ is constant. Let us analyze the structure of supersymmetric extensions  of
 AdS spaces (\ref{ads}). If  $g_{ij}$ is the graded metric tensor
(\ref{gform}) of the AdS space one can define  the following combination of metric tensors
\begin{eqnarray}
\label{gg} 
T_{ijkl}=g_{ik}g_{jl}(-1)^{\epsilon(g)(\epsilon_i+\epsilon_k)+\epsilon_k\epsilon_j}
\end{eqnarray}
which transforms as a tensor field. Therefore a natural generalization of (\ref{ads}) satisfies that
\begin{eqnarray}
\label{sadsR} 
{\cal R}_{ijkl}&=&R(g_{ik}g_{jl}
(-1)^{\epsilon(g)(\epsilon_i+\epsilon_k)+\epsilon_k\epsilon_j}-
g_{il}g_{jk}(-1)^{\epsilon(g)(\epsilon_i+\epsilon_l)+\epsilon_l\epsilon_j+
\epsilon_l\epsilon_k})=\\
\nonumber
&=&(g_{ik}\;R\;g_{jl}
(-1)^{\epsilon_k\epsilon_j}-g_{il}\;R\;g_{jk}(-1)^{\epsilon_l\epsilon_j+
\epsilon_l\epsilon_k})(-1)^{\epsilon(g)},
\end{eqnarray}
where $R$ ($\epsilon(R)=\epsilon(g)$) is a constant. 
  The Ricci tensor satisfies
\begin{eqnarray}
\label{adsRi} 
{\cal R}_{ij}=g^{kl}{\cal R}_{likj}
(-1)^{(\epsilon(g)+1)(\epsilon_k+\epsilon_l)+\epsilon_i\epsilon_k}=R({\cal N}-1)g_{ij}(-1)^{\epsilon(g)}
\end{eqnarray}
and the scalar curvature tensor verifies that
\begin{eqnarray}
\label{adssct} 
{\cal R}=R{\cal N}({\cal N}-1),
\end{eqnarray}
where we  denote
\begin{eqnarray}
\label{adsN}
{\cal N}=\delta^i_i(-1)^{\epsilon_i}
\end{eqnarray}
 and ${\cal N}$ is nothing but the difference between the number of bosonic and
fermionic dimensions of the supermanifold. 

The above Riemannian tensors  obey the following symmetry properties
\begin{eqnarray}
\label{sadsR} 
\nonumber
&&{\cal R}_{ijkl}=-(-1)^{\epsilon_k\epsilon_l}{\cal R}_{ijlk},\quad
{\cal R}_{ijkl}=-(-1)^{\epsilon(g)+\epsilon_i\epsilon_j}{\cal R}_{ijkl},\\
\nonumber
&&{\cal R}_{ijkl}=(-1)^{(\epsilon_i+\epsilon_j)(\epsilon_k+\epsilon_l)}
{\cal R}_{klij}+[1-(-1)^{\epsilon(g)}]g_{il}\;R\;g_{jk}
(-1)^{\epsilon(g)+\epsilon_l(\epsilon_j+\epsilon_k)},\\
\nonumber
&&{\cal R}_{ij}=(-1)^{\epsilon_i\epsilon_j}{\cal R}_{ji}.
\end{eqnarray}
It is  easy to show that in the even case $(\epsilon(g)=0)$ all required symmetry properties for ${\cal R}_{ijkl}$ and ${\cal R}_{ij}$ are satisfied. Therefore the supersymmetric generalization of (\ref{ads}) has the form
\begin{eqnarray}
\label{fadsR}
{\cal R}_{ijkl}=R(g_{ik}g_{jl}
(-1)^{\epsilon_k\epsilon_j}-
g_{il}g_{jk}(-1)^{\epsilon_l\epsilon_j+\epsilon_l\epsilon_k}),\quad
{\cal R}_{ij}=R({\cal N}-1)g_{ij},\quad {\cal R}=R{\cal N}({\cal N}-1).
\end{eqnarray}
In the odd case $(\epsilon(g)=1)$ there exists only one possibility
to satisfy the symmetry requirements: the vanishing of all curvature tensors 
\begin{eqnarray}
\label{odads} 
R=0\longrightarrow {\cal R}_{ijkl}=0,\quad {\cal R}_{ij}=0,\quad
{\cal R}=0.
\end{eqnarray}

\section{Conclusions}

There are two natural geometric structures of  supermanifolds defined by
symmetric and antisymmetric graded tensor fields of the second rank: 
the Poisson bracket defined by an antisymmetric even tensor field of type
$(2,0)$ and the antibracket given by  an symmetrical odd tensor
field of type $(2,0)$.  We have have shown that the geometric structures of  even
and  odd  symplectic supermanifolds  equipped with a symmetric
connection compatible with a given symplectic structure are  very  similar, 
although only in the even  case the scalar curvature has to vanish.
In similar way,  the structures of    even and  odd Riemannian
supermanifolds equipped with the unique symmetric connection
compatible with a given metric structure are also very similar.
However, odd Riemannian supermanifolds are  strongly constrained by the fact that
their scalar curvature has to be constant whereas in the even case the curvature
can have any value.  It is quite remarkable that the strongest restrictions 
on the curvatures arise only  for even symplectic  and odd Riemannian
manifolds. In the case of even Riemannian  or odd symplectic manifolds,
the curvature tensors can be non null and non-constant, respectively. 
There are several practical implications of
the above formal results. The  antisymplectic 
supermanifold  underlying the Batalin-Vilkovisky quantization method  
is just an odd Fedosov supermanifold which as we have shown can have an arbitrary
non-vanishing curvature.  On the other hand, even Riemannian supermanifolds admit 
even AdS superspaces as special case, but there is no analogue  for odd Riemannian supermanifolds, i.e. there are not odd supersymmetric AdS spaces.

\section*{Acknowledgements}
The work of M.A. is partially supported by CICYT (grant FPA2006-2315)
and DGIID-DGA (grant2007-E24/2). P.M.L. acknowledges the MEC for the 
grant (SAB2006-0153). The work of P.M.L. was supported by the RFBR 
grant, project No. 06-02-16346,  the joint RFBR-DFG grant, project No. 06-02-04012, the grant for LRSS, project No. 2553.2008.2.

\end{document}